\documentclass[a4paper,10pt,twoside]{cpc-hepnp}

\usepackage{multicol}
\usepackage{graphicx}
\usepackage{booktabs}
\usepackage[tbtags]{amsmath}
\usepackage{lastpage}
\usepackage{CJK}

\begin{document}

\fancyhead[c]{\small Chinese Physics C~~~Vol. XX, No. XXX (201X)
010201} \fancyfoot[C]{\small 010201-\thepage}

\footnotetext[0]{Received 14 March 2015}
\title{An event mixing technique for Bose-Einstein correlations of two pions in photoproduction around 1 GeV}

\author{
Qing-Hua He$^{1,8}$\email{qhhe@lns.tohoku.ac.jp} 
 \and H. Fujimura$^{1a}$    
 \and H. Fukasawa$^{1}$    
 \and R. Hashimoto$^{1b}$    
 \and Y. Honda$^{1}$    
 \and T. Ishikawa$^{1}$    
 \and T. Iwata$^{2}$    
 \and S. Kaida$^{1}$   
 \and J. Kasagi$^{1}$    
 \and A. Kawano$^{3}$    
 \and S. Kuwasaki$^{1}$    
 \and K. Maeda$^{4}$    
 \and S. Masumoto$^{5}$    
 \and M. Miyabe$^{1}$    
 \and F. Miyahara$^{1c}$    
 \and K. Mochizuki$^{1}$    
 \and N. Muramatsu$^{1}$    
 \and A. Nakamura$^{1}$    
 \and K. Nawa$^{1}$    
 \and S. Ogushi$^{1}$    
 \and Y. Okada$^{1}$    
 \and Y. Onodera$^{1}$    
 \and K. Ozawa$^{6}$ \and Y. Sakamoto$^{3}$ \and M. Sato$^{1}$ \and H. Shimizu$^{1}$ \and H. Sugai$^{1}$ \and K. Suzuki$^{7}$ \and Y. Tajima$^{2}$ \and S. Takahashi$^{1}$ \and Y. Taniguchi$^{1}$ \and Y. Tsuchikawa$^{1}$ \and H. Yamazaki$^{1}$ \and R. Yamazaki$^{1}$  \and H.Y. Yoshida$^{2}$
}

\maketitle

\address{%
$^{1}$Research Center for Electron Photon Science, Tohoku University, Sendai 982-0826, Japan \\
$^{2}$ Department of Physics, Yamagata University, Yamagata 990-8560, Japan\\
$^{3}$ Department of Information Science, Tohoku Gakuin University, Sendai 981-3193, Japan\\
$^{4}$ Department of Physics, Tohoku University, Sendai 980-8578, Japan\\
$^{5}$ Department of Physics, University of Tokyo, Tokyo 113-0033, Japan \\
$^{6}$ Institute of Particle and Nuclear Studies, KEK, Tsukuba 305-0801, Japan\\
$^{7}$ Wakasa-wan Energy Research Center, Tsuruga 914-0192, Japan\\
$^{8}$ Institute of Fluid Physics, China Academy of Engineering Physics, P. O. Box 919-101, Mianyang 621900, China\\
$^{a}$ Present address: Department of Physics, Wakayama Medical University, Wakayama 641-8509, Japan\\
$^{b}$ Present address: Department of Physics, Yamagata University, Yamagata 990-8560, Japan\\
$^{c}$ Present address: Accelerator Laboratory, KEK, Tsukuba 305-0801, Japan\\

}

\begin{abstract}
We have developed an event mixing technique to observe Bose-Einstein correlations (BEC) between two identical neutral pions produced in photo-induced reactions in the non-perturbative QCD energy region. It is found that the missing-mass consistency cut and the pion-energy cut are essential for the event mixing method to effectively extract BEC observables. A Monte Carlo (MC) simulation is used to validate these constraints and confirms the efficiency of this method. Our work paves the way for similar BEC studies at lower energies where the multiplicity of emitted bosons is limited.
\end{abstract}



\footnotetext[0]{\hspace*{-3mm}\raisebox{0.3ex}{$\scriptstyle\copyright$}2013
Chinese Physical Society and the Institute of High Energy Physics
of the Chinese Academy of Sciences and the Institute
of Modern Physics of the Chinese Academy of Sciences and IOP Publishing Ltd}%

\begin{multicols}{2}

\section{Introduction}
The Bose-Einstein correlation (BEC) effect was used for the first time in the 1950's by R. Hanbury-Brown and R. Q. Twiss \cite{Hanbury1954PM45} in astronomy to learn about the size of the photon emission region, i.e. the size of a particular astronomical object that is emitting photons. It is also used as a tool in particle physics to study the space-time properties of boson emitters, especially in particle collision \cite{Alexander2003RPP66, becrefs_pc} and heavy-ion collision \cite{becrefs_hic} reactions with large multiplicity at high energies. 
However, one of the big remaining challenges is to observe BEC effects for low multiplicity reactions at low energies where the observation is strongly obscured by non-BEC factors such as strict kinematical limits and decays of resonances \cite{Klaja2010, Aamodt2011PRD84}. In those reactions, global conservation laws induce significant kinematical correlations between final states particles and complicate the BEC analysis \cite{Chaje2008,He2014}. Poor understanding of the influence of those non-BEC effects limits the effective BEC observation especially for exclusive reactions with only three particles in the final state. A tool to effectively separate BEC effects from those non-BEC correlations is desired. 

Such a method suitable for low multiplicity reactions offers the possibility to determine the spatial and temporal characteristics of the nucleon resonances excited by hadronic or electromagnetic probes in the non-perturbative QCD low energy region. These excited nucleon states might decay back into the ground states, accompanied by emission of final state particles like mesons. Analyzing the BEC effects of the identical mesons will yield information on the duration of the emission process and the size of the excited nucleons. 

The knowledge of the BEC effects of identical particles produced in exclusive reactions with low multiplicity can offer complementary information compared to inclusive reactions at high energies with large multiplicity.   

In general, the wave function of two identical bosons emitted from a source should be symmetric with respect to the exchange of the two bosons (see Fig. \ref{fig15110501}). This symmetrization leads to an enhanced probability of emission if the two bosons have similar momenta. The enhancement is related to the space-time dimensions of the source. Therefore, this effect can be used to study the size of the emitter source. The BEC effects can be measured in terms of the correlation function
\begin{equation}
C_2(1,2)=\frac{P(1,2)}{P(1)P(2)},
\label{eqn13082403}
\end{equation}
where $P(1,2)$ denotes the probability of emitting a pair of identical bosons and $P(i)~(i=1,2)$ is that of emitting each identical boson without BEC. If a sphere of the source with a Gaussian density distribution is assumed, Eq. (\ref{eqn13082403}) can be expressed in an analytic form \cite{Alexander2003RPP66}
\begin{equation}
C_2(Q)=N(1+\lambda_2e^{-r_{0}^{2}Q^2 }),
\label{eqn13082412}
\end{equation}
where $N$ is the normalization factor, $Q$ denotes the relative momentum defined by $Q^2=-(p_1-p_2)^2=M^2-4\mu^2$ (M is the invariant mass of the two identical bosons of mass $\mu$ with the four-momentum $p_i$), and $r_{0}$ denotes a measure of the source dimension. The parameter $\lambda_2$ reflects the BEC strength for incoherent boson emission \cite{Deutschmann1982} and varies from 0 (completely coherent case) to 1 (totally chaotic limit). Experimental factors such as particle misidentification and detecting resolution inevitably affect the measurement of bosons' correlations and consequently are reflected in $\lambda_2$. Equation (\ref{eqn13082412}) indicates the BEC effect would be observed as an enhancement at low $Q$.

\begin{center}
\includegraphics[width=0.95\linewidth]{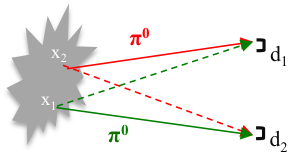}
\figcaption{\label{fig15110501} (color online) Schematic illustration of Bose-Einstein correlations between two identical particles. These two particles are emitted from points $x_1$ and $x_2$ in a source.  They are detected at two separate positions $d_1$ and $d_2$ with momenta $p_1$ and $p_2$, respectively. Two possible trajectories (full and dashed lines) cannot be distinguished. In the case of hadrons, the distance between the emission points is of the order of a femtometer. The distance between the detector and the particles emission source is of the order of a meter.}
\end{center}

The BEC effect is investigated by studying the spectra of their relative momentum $Q$. The correlation is quantified by the correlation function (\ref{eqn13082403}) which is defined as the ratio of the $Q$ distribution of real data to that of reference samples free from BEC effects. A set of valid reference samples should be identical to the real data in all aspects but free of BEC effects. It is therefore a big challenge in a BEC study to find the appropriate reference samples, which affect the BEC result rather strongly and are thus the main source for the overall systematic errors. One of the methods to make a reference sample is the event mixing method that generates an artificial event by taking two particles from different real events\cite{Kopylov1974}, as illustrated schematically in Fig. \ref{fig15062001}.  

Event mixing techniques work well, for example, in estimating the background of a resonance and in observing BEC effects, both for inclusive reactions with large multiplicity. In 1984, D. Drijard, H.G. Fischer and T. Nakada \cite{Drijard1984NIMA225} studied in detail the application of event mixing in the extraction of resonance signals in inclusive reactions. The features of the invariant mass distribution obtained from combinations of particles from different events, i.e. event mixing, was investigated. They concluded that event mixing can reproduce the shape of the uncorrelated invariant mass distribution and be used to observe resonance signals via subtracting the uncorrelated invariant mass distribution from the original one. 

Event mixing is also used as a tool to observe BECs for identical boson pairs produced in inclusive reactions with large multiplicities, in which the kinematical limits induced by global conservation laws are relatively weak and thus its influence on the BEC analysis can be ignored. However, in the low-multiplicity case, global conservation laws yield pronounced N-body correlations and consequently make BEC analysis obscure. Several earlier works investigated the features of these correlations and tried to establish an analytic formalism to account for these correlations \cite{Bertsch1994, Utyuzh2007, Chaje2008}. Although their works improve our understanding of these correlations, the corresponding method for solving the problem is missing. As for exclusive reactions especially with only three final state particles, global conservation laws impose strong correlations on the final state particles. It is a remaining challenge to find an effective event mixing technique which not only can preserve the phase space correlations but also eliminate the other kind of correlations like resonances and BEC effects. 


In this work we study the features of event mixing for three body reactions by taking double pion photoproduction as an example. We aim to develop an event mixing method which is capable of identifying BEC effects and of extracting correct BEC parameters $r_0$ and $\lambda_2$ in the low energy region where only a few mesons are produced simultaneously. We take the reaction $\gamma p \rightarrow \pi^{0}  \pi^{0} p$ as an example to demonstrate this technique. And an extended version is also developed to make it applicable to a more general reaction $\gamma p \rightarrow \pi^{0}  \pi^{0} X$. Although developed on the basis of double neutral pion photoproduction, this event mixing method can be used for similar BEC studies of the three-body reactions which emit only two bosons at low energies.  

\end{multicols}
\begin{center}
\includegraphics[width=0.95\linewidth]{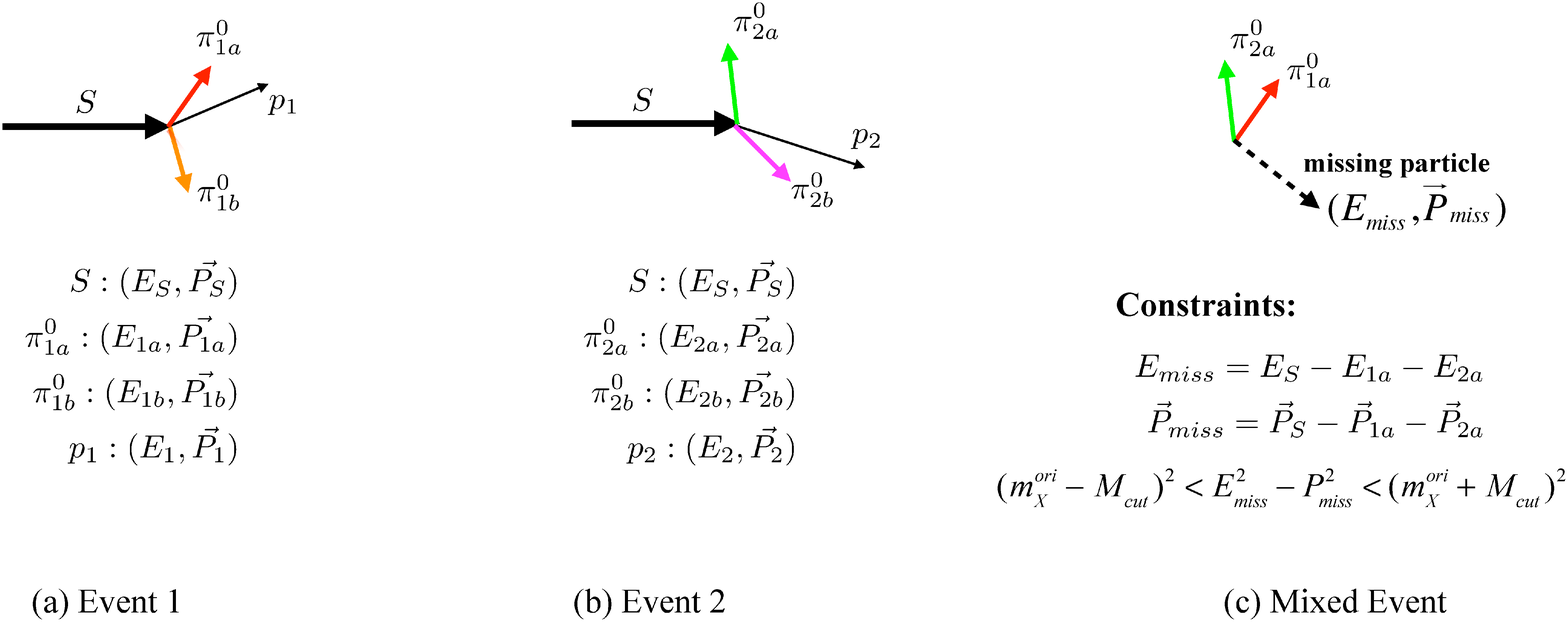}
\figcaption{\label{fig15062001}  (color online) Mixed events (c) are generated by taking two pions from event 1 (a) and event 2 (b), one by one. Here, the reaction $\gamma p \rightarrow \pi^{0}  \pi^{0} p $ is taken as an example. In constructing mixed events (c), the missing-mass consistency cut governs the mixing process. The $m_X^{ori}$ stands for the missing mass for either of the original events and its value is equal to the proton mass in this example.}
\end{center}
\begin{multicols}{2}

\section{Event mixing method}

An event mixing method is used to eliminate BEC effects in a reference sample via constructing a `mixed-event' by taking two identical bosons from different events. This method, however, eliminates all other kinds of correlations, like those due to kinematic conservation laws. Additional appropriate kinematic cuts are, therefore, needed to preserve these correlations not arising from BEC. Here we use the reaction with two neutral pion production to demonstrate these cuts. In order to make a valid reference sample, the event mixing method should contain the following cut conditions: a) missing-mass consistency cut; b) pion-energy cut. 

\subsection{Missing-mass consistency cut}
The missing-mass consistency cut is used to ensure that mixed events satisfy four-momentum conservation. It requires that the mass of a missing particle $X$ in the mixed event $\pi^{0}  \pi^{0} X$ should satisfy the relation: $|m_{X}^{mix}-m_X^{ori}|<M_{cut}$, where $m_{X}^{mix}$ and $m_{X}^{ori}$ are the missing particle mass for the mixed event and that for the original event, respectively. The prerequisite of this cut is that the values of $m_{X}^{ori}$ for the two original events should be close enough as described in the sample dividing later. The four-momentum of the missing particle $X$ is given by the four-momentum conservation of the reaction: $P_{X}=P_{beam}+P_{target}-P_{\pi1}-P_{\pi2}$ where $P_{\pi1}$ and $P_{\pi2}$ are the four-momenta of two mixed pions. The cut window, $M_{cut}$, is generally set to be 10 MeV.

A mixed event is accepted if the missing mass of the generated event meets the constraint, as indicated in Fig. \ref{fig15062001} (c), where the reaction $\gamma p \rightarrow \pi^{0}  \pi^{0} p $ is taken as an example. To use this cut properly, the two events to be mixed must be in the same energy bin of incident photons (the bin size = 20 MeV). The incident photons are generated so as to have a $1/E_{\gamma}$ intensity distribution, where $E_{\gamma}$ denotes the incident photon energy.

A Monte Carlo simulation based on the $\gamma p \rightarrow \pi^{0}  \pi^{0} p$ reaction shows the missing mass cut can successfully reject unphysical mixed events in the large Q region, as illustrated in Fig. \ref{fig14010301} (a). The correlation functions, calculated as the pure phase space $Q$ distribution of $\pi^{0}  \pi^{0}$ normalized to two reference spectra obtained by event mixing with and without the missing-mass consistency cut, indicate this cut can yield a flatter correlation function compared to the case without any cut condtion. In the simulation, the $\gamma p \rightarrow \pi^{0}  \pi^{0} p$ events are generated randomly following a pure phase space distribution. The phase space events are generated by using a ROOT utility named ``TGenPhaseSpace`` \cite{TGenPhaseSpace}. The code is based on the GENBOD function, implemented in the CERN library, using the Raubold and Lynch method \cite{James1968}. To generate a certain type of event randomly, a total energy ($E_{tot}$), total multiplicity (N) and a list of masses ($m_i$) of emitted particles are requested as input parameters. For each event, it returns a weight proportional to the probability of natural appearing. This weight is based upon the phase-space integral $R_N$
 \begin{equation}
R_N=\int^{4N}\delta^4(P-\sum_{j=1}^{N}p_j)\prod\limits_{i=1}^N\delta(p_i^2-m_i^2)d^4p_i,
\label{eqn16032101}
\end{equation}
where $P$ and $p_i$ are the total four momentum of the whole system and that of individual emitted particles, respectively. The uniform event distribution in the Dalitz plot, a two dimensional plot of $m(p, \pi)$ versus $m(\pi, \pi)$ as shown in Fig. \ref{fig14010301} (b), shows the event generation is performed properly. 

\begin{center}
\includegraphics[width=0.49\linewidth]{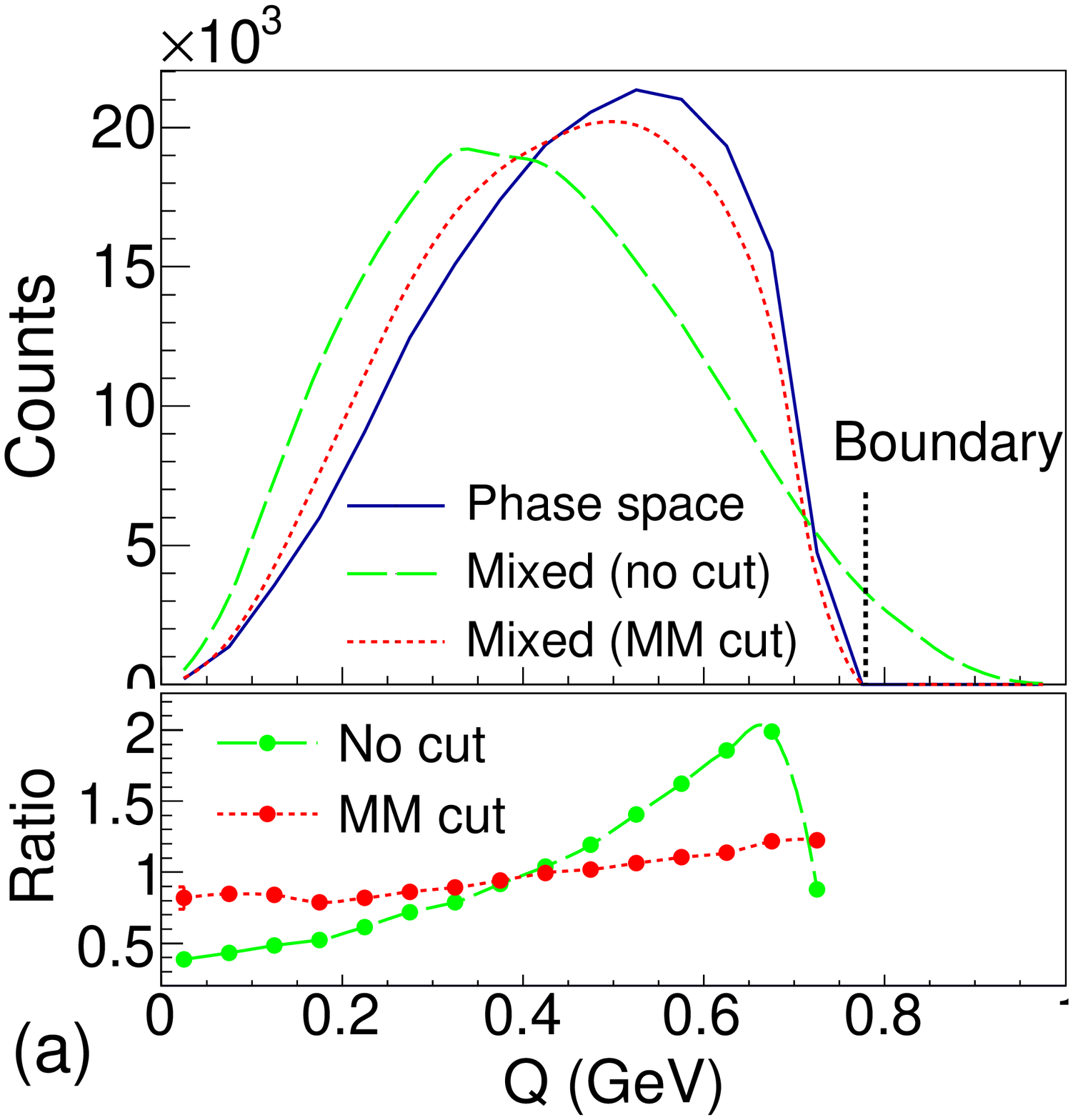}~~~
\includegraphics[width=0.50\linewidth]{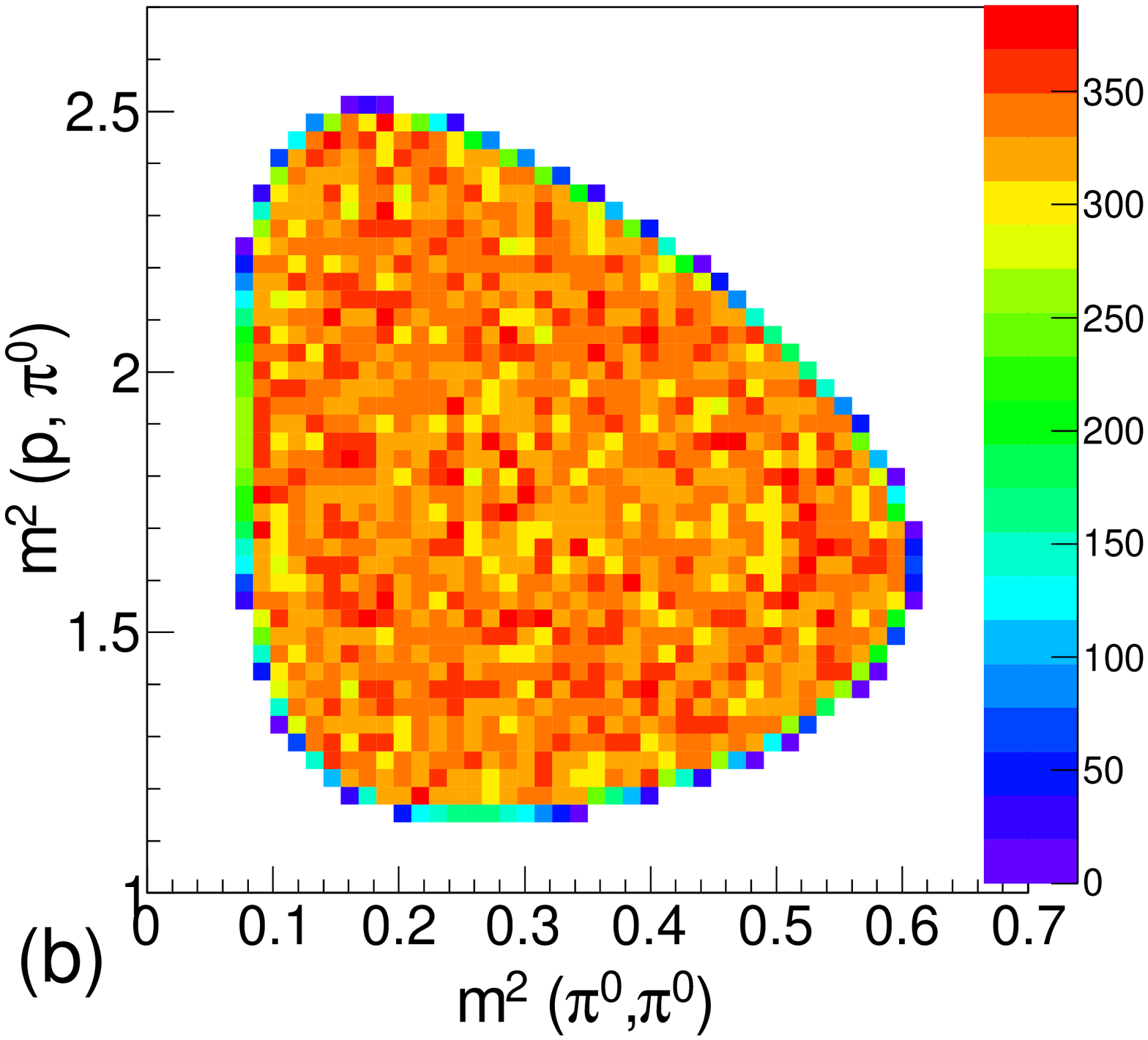}
\figcaption{\label{fig14010301} (color online) (a) $Q$ distributions of two pions for mixed events. The solid line indicates phase space events, and the dashed (long dashed) line represents mixed events with (without) the missing-mass (MM) consistency cut. It shows the missing-mass consistency successfully rejects unphysical events produced by event mixing. The ratios of the phase space $Q$ spectrum to the event-mixing-obtained ones with and without the MM consistency cut are also shown in the lower panel respectively, indicating the MM consistency cut is still unable to make a flat correlation function though it improves a lot. (b) Dalitz plot of the phase space $\gamma p \rightarrow \pi^{0}  \pi^{0} p $  events generated by a Monte Carlo simulation. }
\end{center}

Fig. \ref{fig14050605} gives typical distributions of the missing mass of $\pi^0\pi^0$ for the reactions $\gamma p \to \pi^0\pi^0p$ and $\gamma p \rightarrow \pi^0\pi^0X$, obtained in a real experiment with an incident photon beam of about
1 GeV. The first peak in Fig. \ref{fig14050605} (b) corresponds to the proton, while the second wide peak reflects other contributions like $\gamma p \to \eta p \to \pi^0\pi^0\pi^0p$, $\gamma p \rightarrow \pi^0\pi^0\pi^+n$, etc. We assume the missing mass spectrum can be given with gaussian functions in the simulation to meet each experimental situation. A simulated missing mass spectrum is divided into several bins, the size of which is set to be 10 MeV in the present study. In the event mixing, two original events must be chosen from the same bin of the original missing mass spectrum, yielding two mixed events having usually different missing masses. The event mixing continues until the whole missing mass spectrum for the mixed events becomes that for the original events. This is indeed useful for the inclusive case, because the mass of the missing particle $X$ widely varies especially at high energy bins.

\begin{center}
\includegraphics[width=0.45\linewidth]{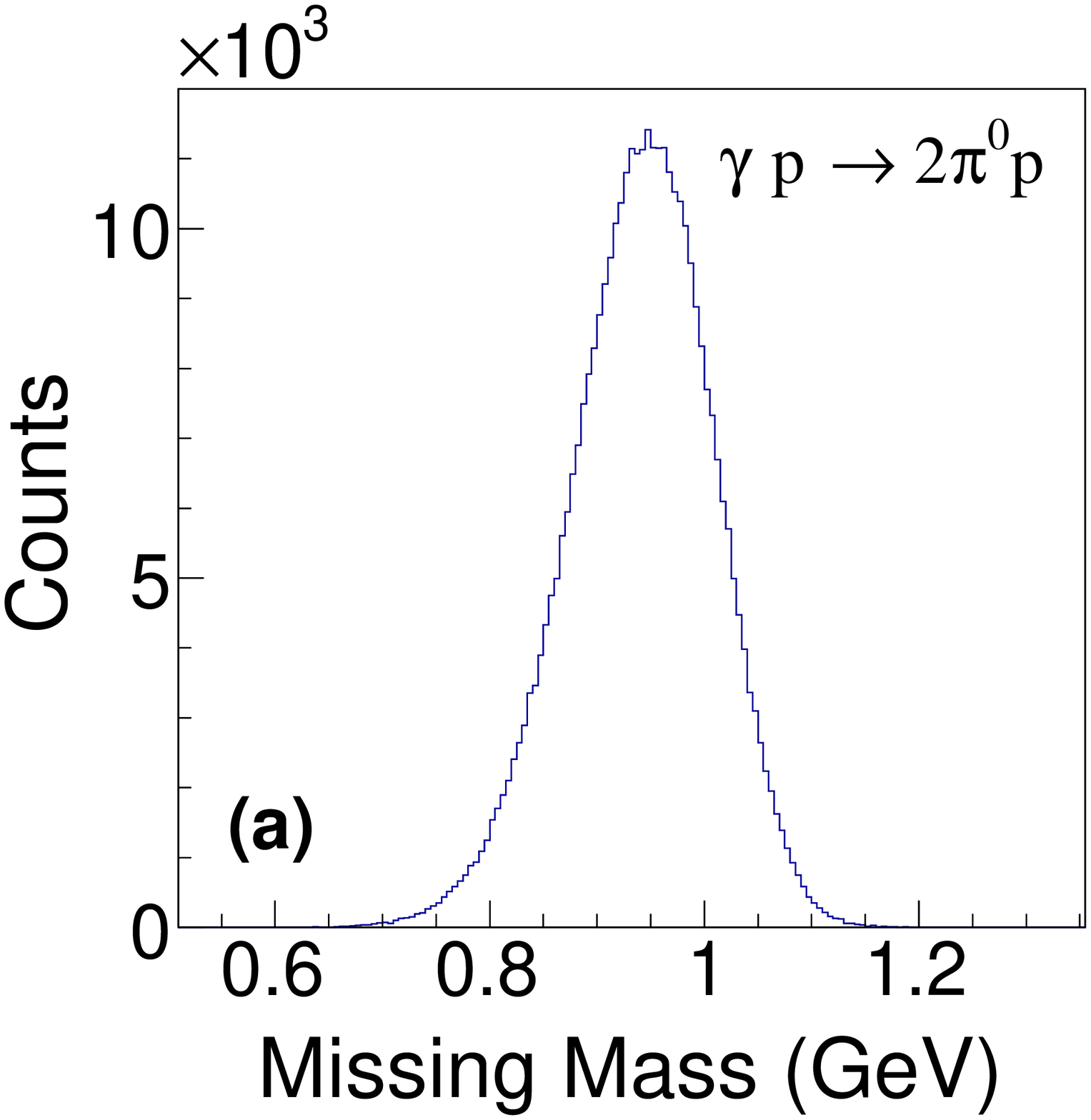}~~~
\includegraphics[width=0.45\linewidth]{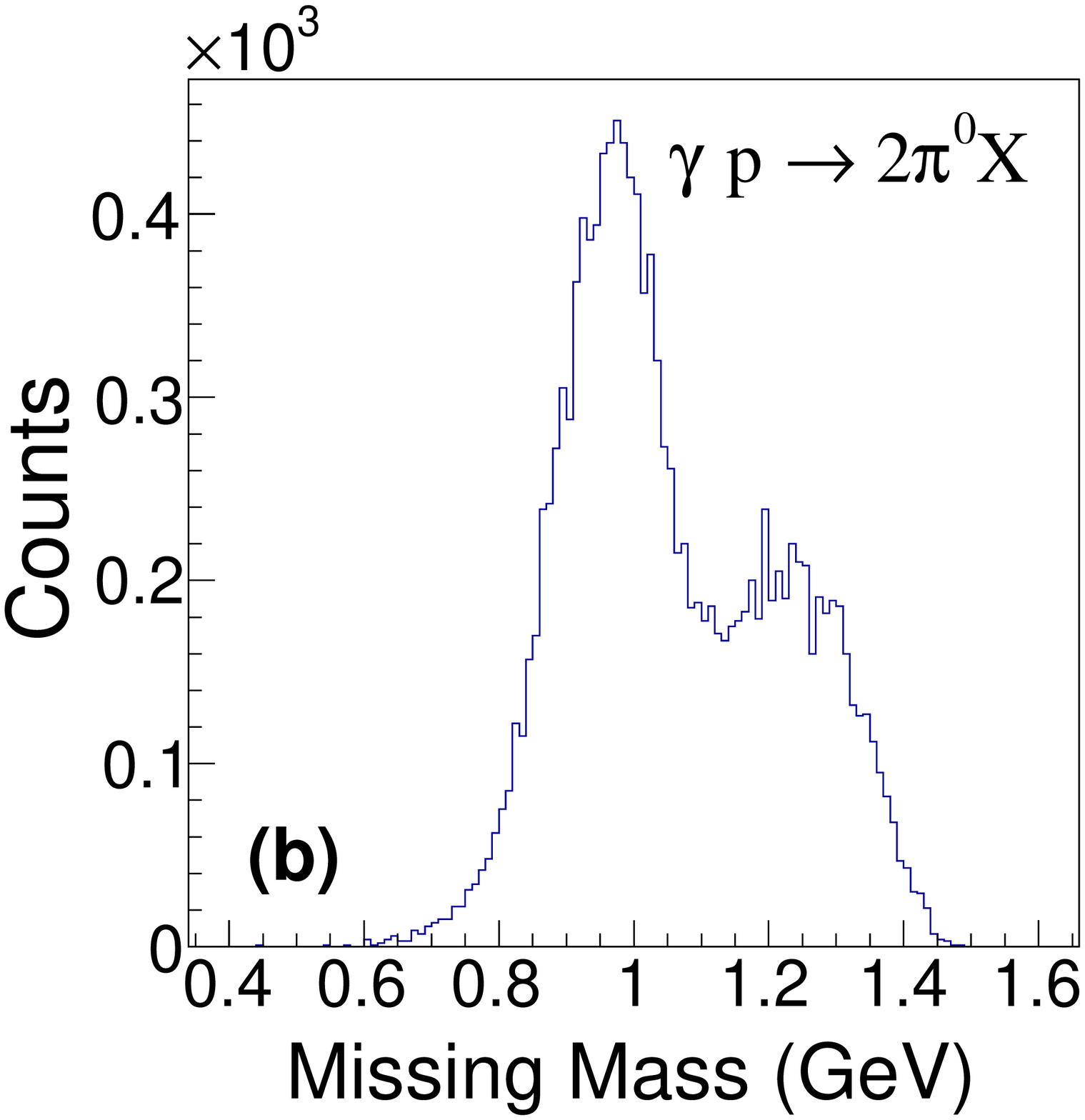}
\figcaption{\label{fig14050605} Typical missing mass spectra of $\pi^0\pi^0$ for (a) the reactions $\gamma p \to \pi^0\pi^0p$ with an incident photon beam energy in $0.74\le E_{\gamma} \le 1.15$ GeV and for (b) the inclusive reaction $\gamma p \rightarrow \pi^0\pi^0X$ in $1.10 \le E_{\gamma} \le 1.15$ GeV, obtained in a real experiment.}
\end{center}

\subsection{Pion-energy cut}
A desirable event mixing method should make the correlation function flat in the whole Q region for the pure phase space events of $\gamma p \to \pi^0 \pi^0 p$ without BEC effects. Although the missing-mass consistency cut guarantees that the mixed events are confined in the allowed region, it fails to make a Q distribution of the mixed events identical to that of the original events. Only the missing-mass consistency cut is not able to make a flat correlation function and another cut is needed. Many variables, like the azimuthal angle and polar angle of pion, are investigated, and finally it is found that the pion energy is sensitive to the slope of the correlation function, which is an indicator of the degree of being flat. A pion-energy cut is therefore developed empirically on the basis of the pure phase space events of $\gamma p \to \pi^0 \pi^0 p$. It requires that both pions should satisfy the condition $E_{\pi}<E_{\pi}^{max}$, where $E_{\pi}^{max}$ is optimized as described below. The events not passing the cut condition are rejected from the event mixing. This energy cut removes the events near the phase space boundary. 

In order to find the optimum pion-energy cut condition, several different values of $E_{\pi}^{max}$ are tested in the energy range from $0.3E_{\gamma}$ to $0.7E_{\gamma}$ with a step size of $0.01E_{\gamma}$, where $E_{\gamma}$ denotes the incident photon energy. Fig. \ref{fig14050101} shows typical correlation functions obtained in the simulations with a different pion-energy cut $E_{\pi}^{max}$ ranging from $0.45E_{\gamma}$ to $0.55E_{\gamma}$. A linear function $y=p0+p1\cdot Q$ is fitted to each correlation function, which indicates the ratio of the $Q$ distribution of the generated events to that of the mixed events. The best cut condition should yield the slope parameter $p1=0$, so that the optimum value of $E_\pi^{max}$ is determined based on a polynomial fitting for the plot of p1 with the different pion-energy cuts, as shown in Fig. \ref{fig14050102}.

\begin{center}
\includegraphics[width=0.99\linewidth]{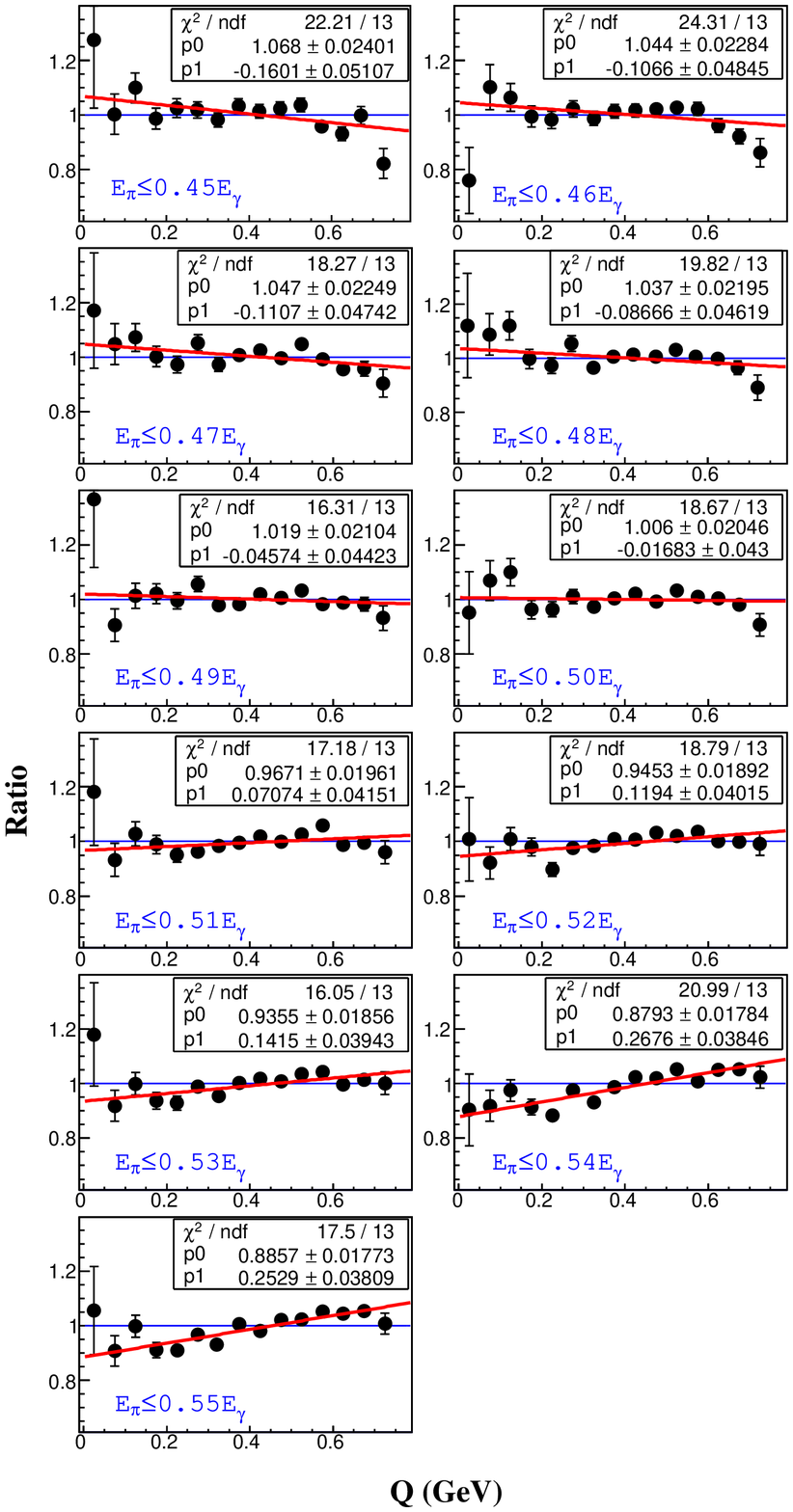}
\figcaption{\label{fig14050101} The ratio of the $Q$ distribution of phase-space Monte Carlo events to that of mixed events. The missing-mass consistency cut and different pion-energy cuts are applied to get the mixed events. The pion-energy cut condition is indicated by the text like `$E_{\pi}\le xE_{\gamma}$' in each panel. A function $y=p0+p1\cdot Q$ is fitted to each ratio with fitting parameters shown in the same panel. }
\end{center}

\begin{center}
\includegraphics[width=0.9\linewidth]{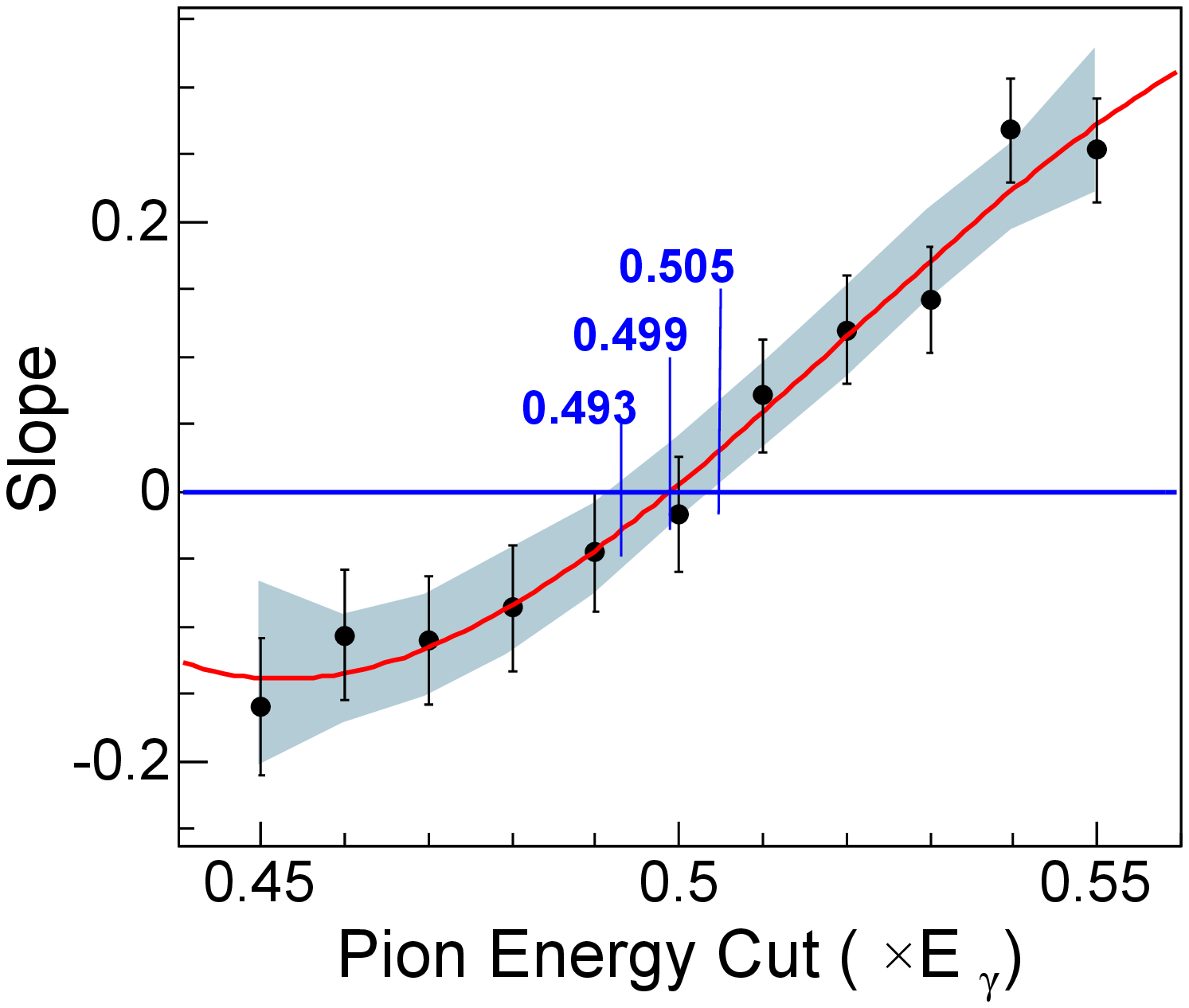}
\figcaption{\label{fig14050102} (color online) The slope parameter $p1$ as a function of the pion-energy cut for the $\gamma p \rightarrow \pi^0\pi^0p$ reaction. The fitted polynomial function (red solid curve) intersects the line of p1=0 at $E_\pi^{max} = 0.499 E_\gamma$, as indicated by the text `0.499'. The vertical lines at `0.493' and `0.505' correspond to the boundaries of the fitting uncertainty (filled region) with $95\%$ confidence interval.}
\end{center}

The pion energy spectrum of the original and the mixed events shows that the pions of lower energy are easier to be mixed  than of higher energy (see Fig. \ref{fig16011102} (a)). Pion energy cut rejects higher energy pions in the event mixing and therefore eliminates the effect of this bias on the Q distribution of the mixed events. This can also be illustrated in a two dimensional plot of two pions' energies, as shown in Fig. \ref{fig16011102} (b). It shows that high energy pion must be coupled with a low energy pion due to the energy conservation. But the pions in the squared area ($E_{\pi_1}<0.575$, $E_{\pi_2}<0.575$ GeV ) have no heavy energy dependence when finding a pion to construct a mixed events. 

\begin{center}
\includegraphics[width=0.47\linewidth]{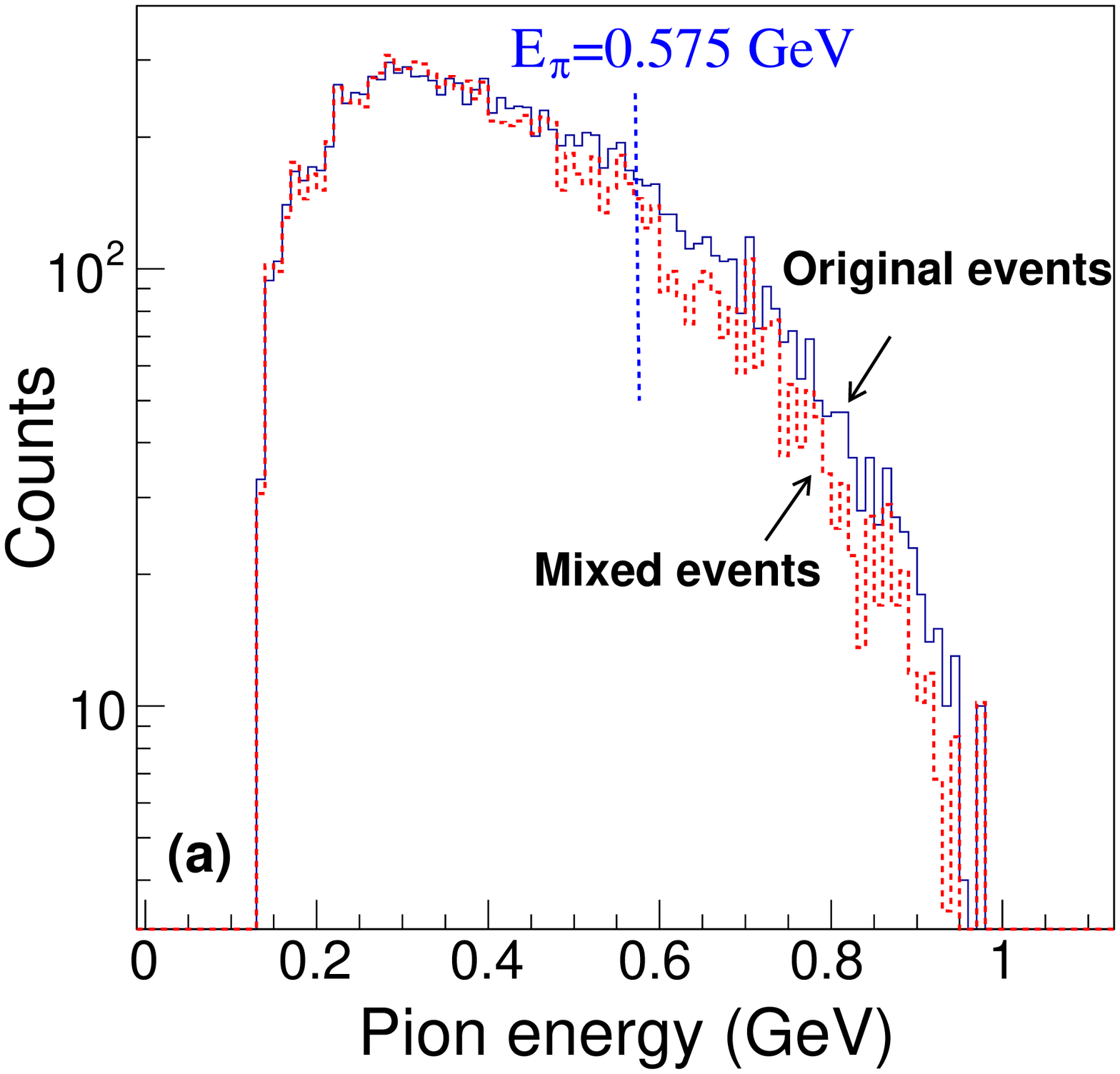}
\includegraphics[width=0.50\linewidth]{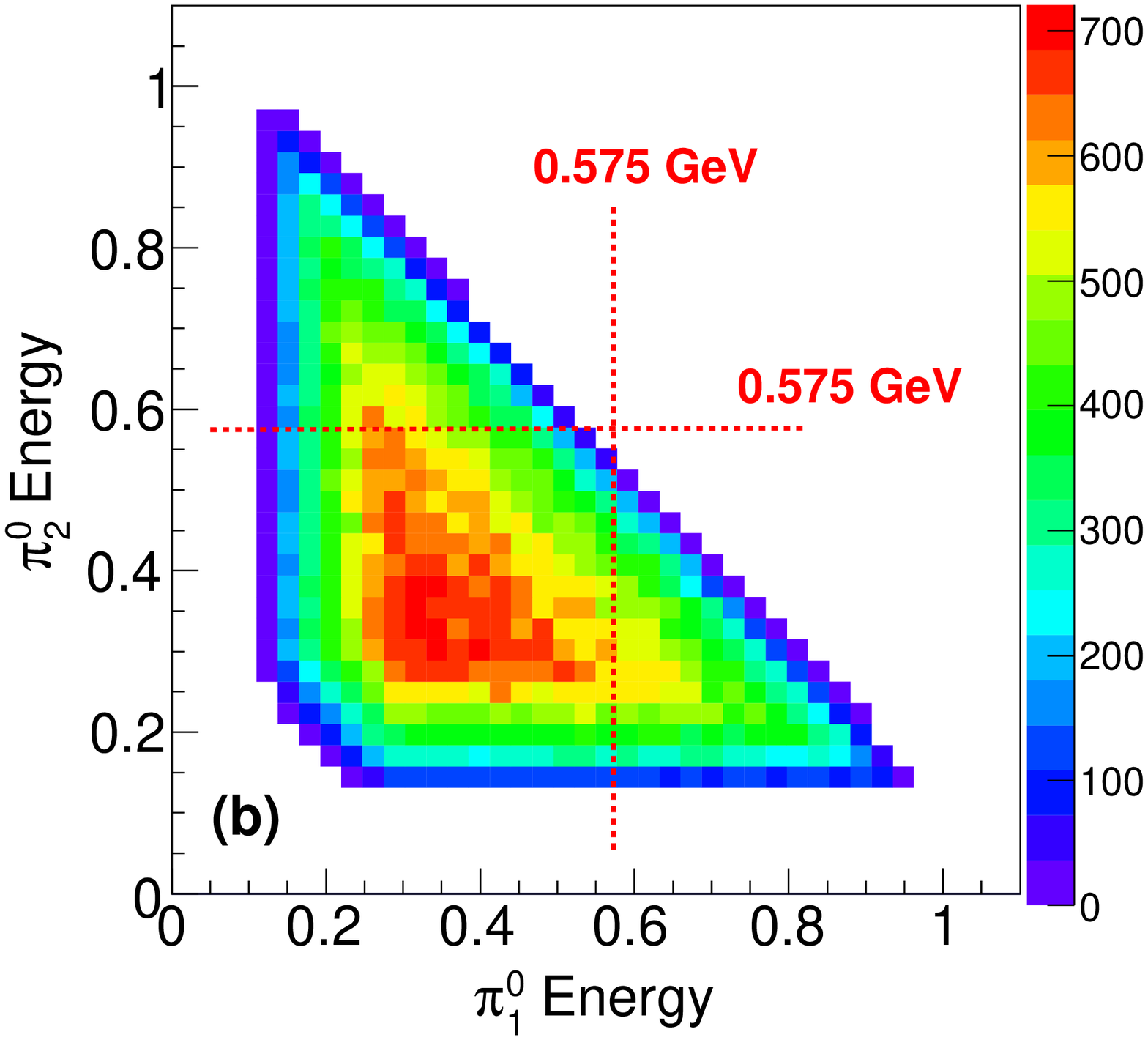}
\figcaption{\label{fig16011102} (color online)  (a) Pion energy spectra of the mixed/original events and (b) two dimensional plot of two pions' energies for the $\gamma p \rightarrow \pi^0\pi^0p$ reaction at the indecent photon energy of $E_{\gamma}=$ 1.15 GeV. The mixed events in the plot are obtained with the missing-mass consistency cut.}
\end{center}

For both the inclusive reaction $\gamma p \rightarrow \pi^0\pi^0X$ and the exclusive reaction $\gamma p \rightarrow \pi^0\pi^0p$ , the maximum energy $E_{\pi}^{max}$ is determined for each missing mass bin $m_{X}$. The optimum pion-energy cut is found to depend on $m_{X}$. In the simulation, we produce nine samples of the pure phase space process $\gamma p \rightarrow \pi^0\pi^0X$ corresponding to the nine $m_{X}$ points varying from $m_p-100$ MeV to $m_p+300$ MeV with a step size of 50 MeV. Finally, the $m_{X}$ dependent pion-energy cut $E_{\pi}^{max}$ is obtained as $E_{\pi}^{max}/E_{\gamma}=1.1-0.64m_{X}$, where $m_{X}$ is given in the unit of GeV. For a certain $m_{X}$ and $E_{\gamma}$, events with either pion's energy greater than $E_{\pi}^{max}$ are rejected from the event mixing. Fig. \ref{fig14050604} shows the best pion-energy cut dependence on the missing particle mass $m_{X}$ at the incident photon energy $E_{\gamma}=1.1$ GeV.

\begin{center}
\includegraphics[width=0.9\linewidth]{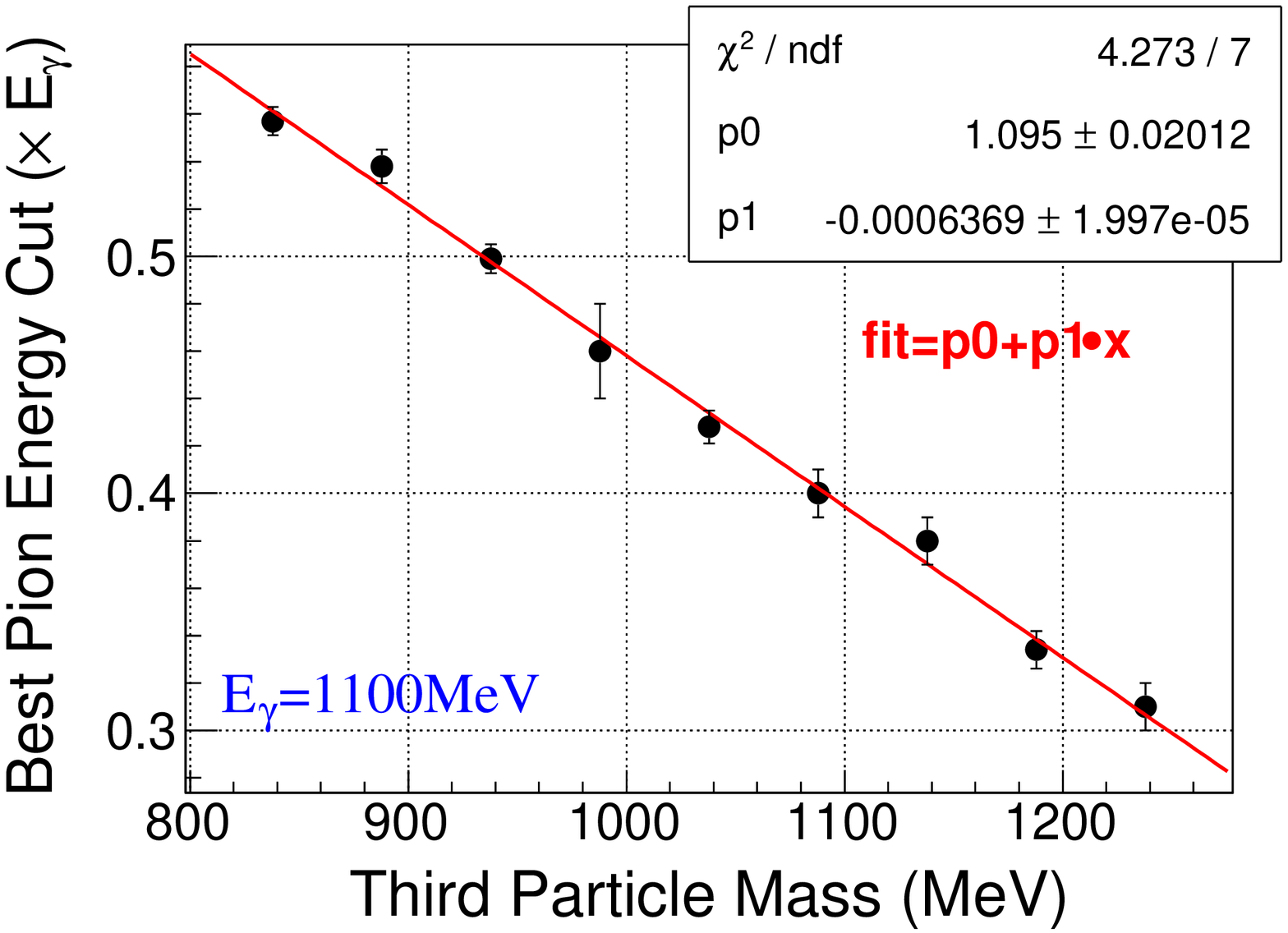}
\figcaption{\label{fig14050604} The optimum pion-energy cut as a function of the missing mass $m_X$ of inclusive reaction $\gamma p \rightarrow \pi^0\pi^0X$ at the incident photon energy of 1100 MeV.}
\end{center}

\section{Verification of event mixing method }
\label{sec_testMixingMethod}
We performed several numerical tests to demonstrate the efficiency and correctness of the event mixing method. Without loss of generality, we adopt the MC simulations of the reaction $\gamma p \rightarrow \pi^0\pi^0X$ where $X$ represents the missing particle/s. The missing mass spectrum $m_X$ is adjusted so that it is similar to that in Fig. \ref{fig14050605} (b). This is done by combining two Gaussian distributions together with a certain weight for each one. The first peak of the missing mass is represented as a Gaussian function with $\mu$ = 938 MeV, $\sigma$ = 70 MeV and a weight of 0.7, and the second peak with $\mu$ = 1200 MeV, $\sigma$ = 70 MeV and a weight of 0.3 (These weights are estimated roughly by the areas under the proton peak and that under the residual area in the missing mass distribution). 

We generate eight sets of samples with BEC effects and one sample set free of the BEC effects (called `noBEC'). 
Table \ref{tbl14052001} lists the different combinations of the BEC parameters $r_0$ and $\lambda_{2}$ at an incident photon energy $E_\gamma = 1.1$ GeV. The BEC effects are taken into account through Eq. (\ref{eqn13082412}). At first a mass $m_X$ is generated according to the missing mass distribution. Then a $\pi^0\pi^0X$ event is generated for the mass $m_X$, giving a $Q$. Each set of BEC samples is produced with a certain $r_0$ and a $\lambda_2$ by filtering a pure phase space event sample with a weight calculated by Eq. (\ref{eqn13082412}) in terms of $Q$.

\begin{center}
\tabcaption{\label{tbl14052001} Input BEC parameters $r_0$ and $\lambda_2$ for simulation samples of the inclusive reaction $\gamma p \to \pi^0\pi^0X$ at $E_\gamma$=1.1 GeV.}
\begin{tabular}{lll}
\hline
\hline
 Event Sample  & $r_0$ (fm) & $\lambda_2$  \\
\hline
noBEC  & $--$      & $--$   \\
BEC,1   & $0.4$   & 1.0   \\
BEC,2   & $0.7$   & 1.0   \\
BEC,3   & $1.0$   & 1.0   \\
BEC,4   & $1.3$   & 1.0   \\
BEC,5   & $0.4$   & 0.5   \\
BEC,6   & $0.7$   & 0.5   \\
BEC,7   & $1.0$   & 0.5   \\
BEC,8   & $1.3$   & 0.5   \\
\hline
\end{tabular}
\end{center}

Although the pion energy cut inevitably reshapes the sample statistics, it is an effective cut if it does not change the BEC induced statistics as long as the BEC analysis is concerned. The consistent correlation functions from the whole sample and the partial sample obtained by the pion energy cut $E_{\pi}<0.5E_{\gamma}$ shows the pion energy cut does not change the BEC statistics (see Fig. \ref{fig16031601}).

\begin{center}
\includegraphics[width=0.9\linewidth]{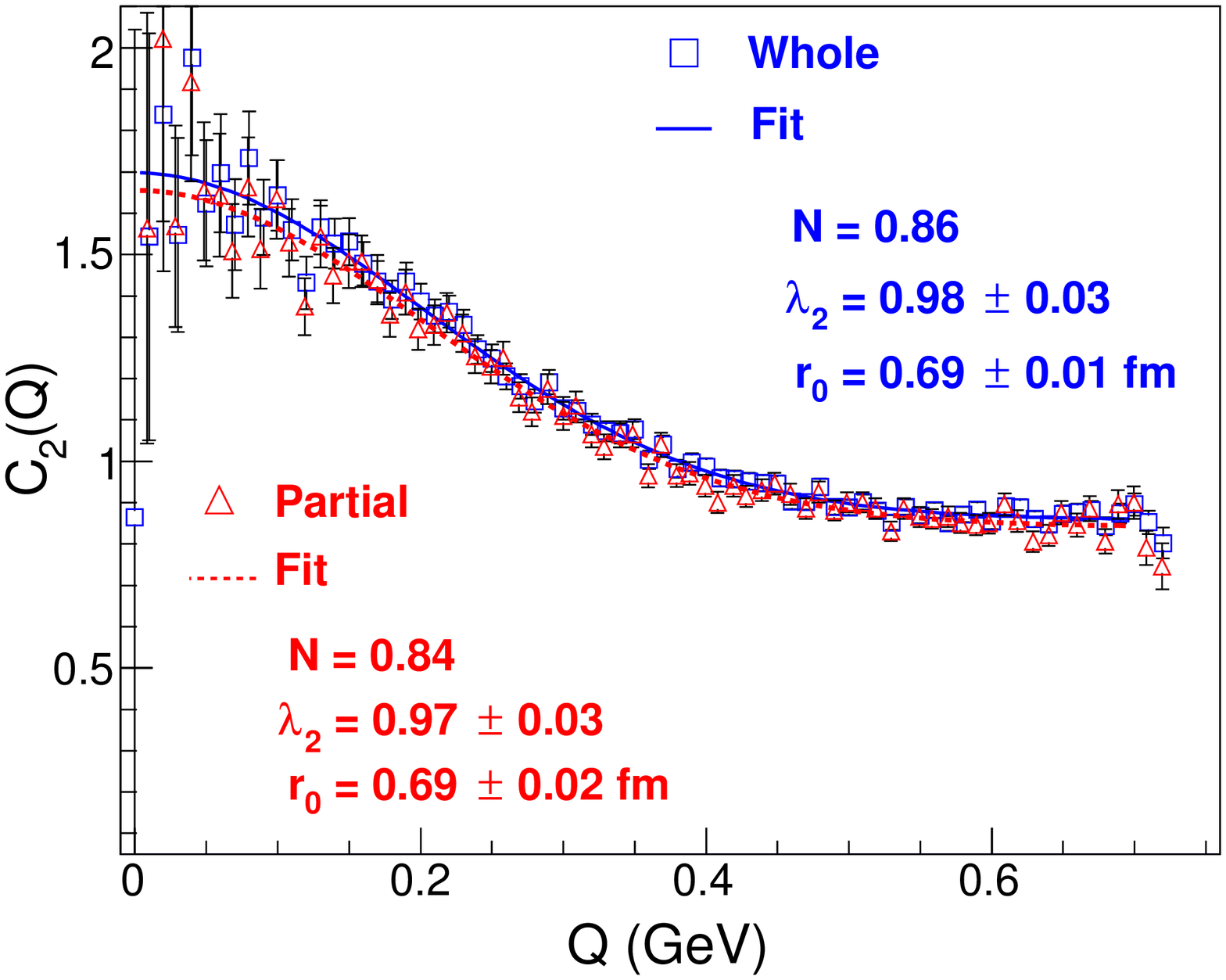}
\figcaption{\label{fig16031601} Two typical correlation functions obtained as the ratio of $Q$ spectrum of `BEC,2' sample to that of `noBEC' sample, with whole events and partial events survived the pion energy cut $E_{\pi}<0.5E_{\gamma}$, respectively.}
\end{center}

The developed event mixing method with the missing-mass consistency cut and the pion-energy cut is employed for the generated samples. The masses of the missing particles both in the mixed events and the original events are shown in Fig. \ref{fig14052003}. A good agreement is found between them, demonstrating the efficiency of the strategy that we only mix two events in the same $m_X$ bin.

\begin{center}
\includegraphics[width=0.99\linewidth]{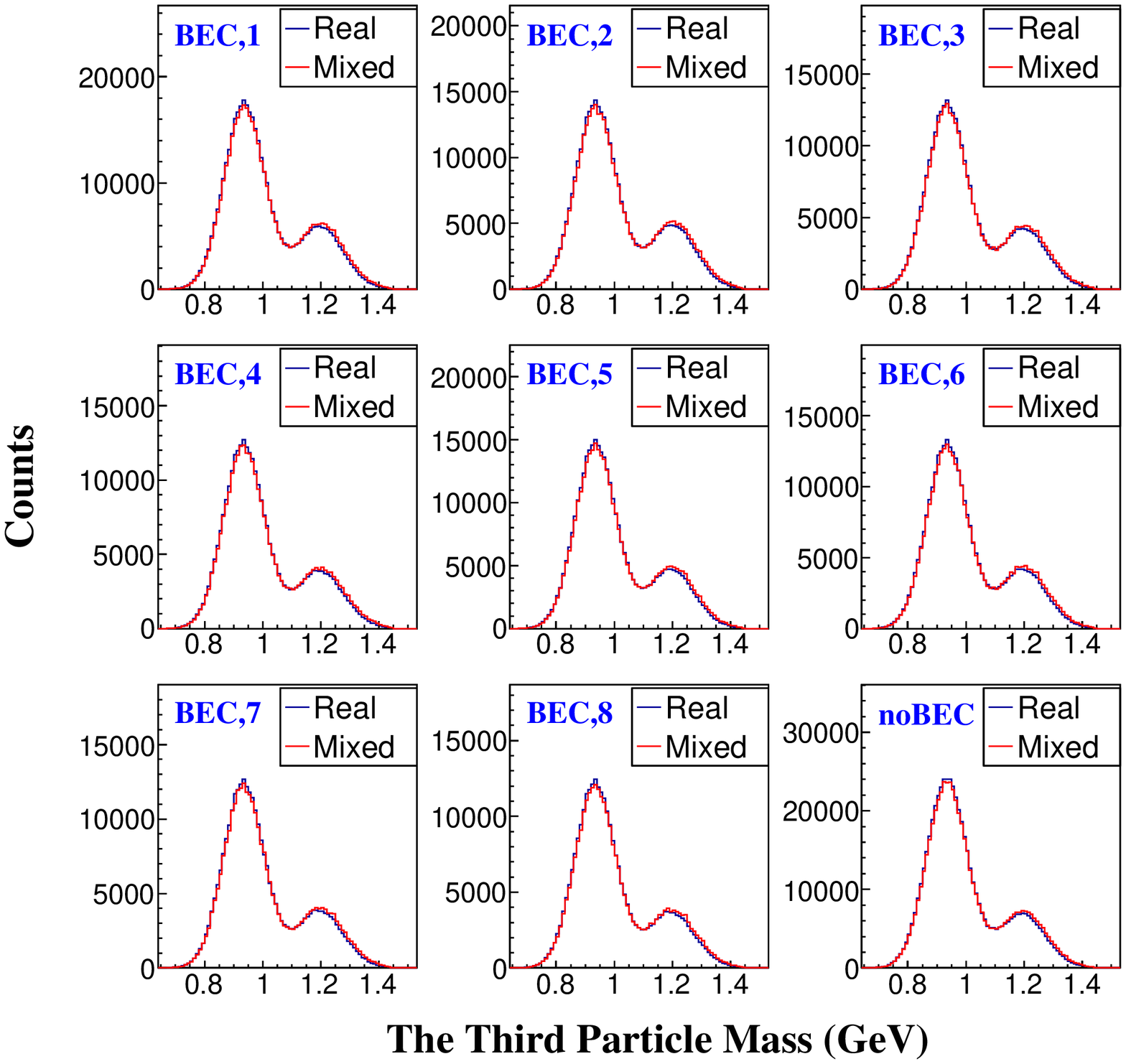}
\figcaption{\label{fig14052003} Missing masses for the mixed
events from the $\gamma p \to \pi^0 \pi^0 X$ simulation. For comparison, the missing masses of the original events are shown as well. }
\end{center}

The corresponding correlation functions (the ratio of the $Q$ distribution of BEC samples to that obtained by event mixing) are plotted with filled circles in Fig. \ref{fig15062002}. For comparison, the ratio between the $Q$ distribution of each set of BEC samples and that of `noBEC' samples is also shown by open triangles. In the calculation, the $Q$ distributions are normalized to the integral over the region 0.6 GeV< $Q$ < 0.65 GeV. For the `noBEC' samples, the event mixing method makes a flat distribution of the correlation function, as expected. For the BEC samples, we clearly observe the BEC effects, indicating this method can be used to observe BEC effects for the inclusive reaction $\gamma p \rightarrow \pi^0\pi^0X$. 

Eq. (\ref{eqn13082412}) is fitted to the two types of correlation functions at each panel to get the values of $r_0$ and $\lambda_2$. By comparing the fit result from event mixing with those from pure phase space events, it is found this method overestimates $r_0$ and underestimates $\lambda_2$ in most cases except for the tests with the smallest $r_0$. Typical uncertainties of $r_0$ and $\lambda_2$ are $12\%$ and $20\%$ respectively when input $r_0=1.0$ fm and $\lambda_2=1.0$. And the $\chi^2/ndf$ values of the fit to the correlation functions obtained with the mixed events are worse than those obtained with the phase space events. These results indicate some flaws in this event mixing method reduce the capability of the BEC parameters observing. The non-flat structure in the high Q region can also demonstrate the limitation of this method. Those flaws are the main source of the systematic error in making use of this mixing method. In the practical use, the systematic errors should be evaluated by using a similar simulation as the original events.

\begin{center}
\includegraphics[width=0.99\linewidth]{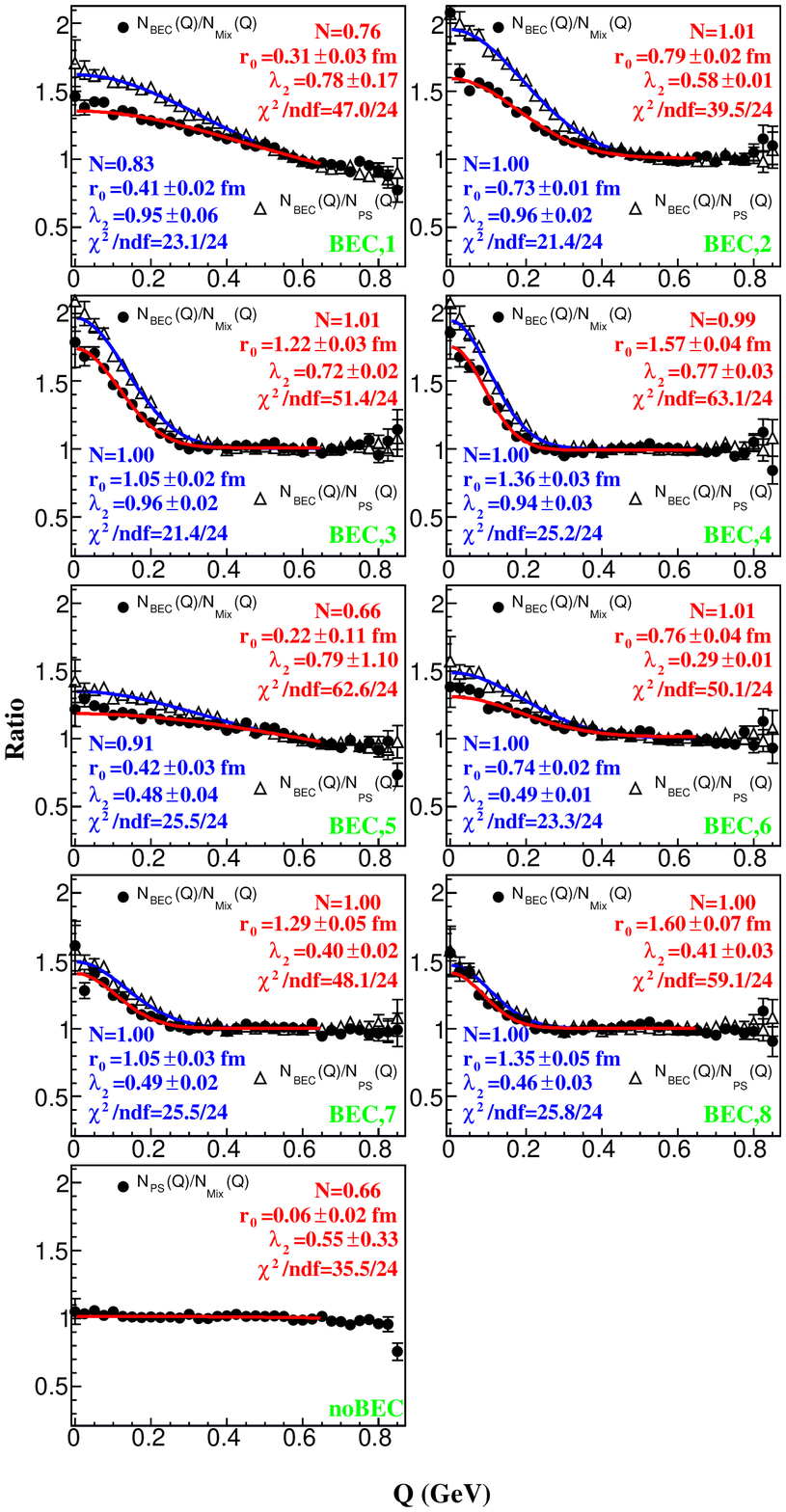}
\figcaption{\label{fig15062002} The ratio of the $Q$ distribution of the generated BEC/noBEC sample, $N_{BEC}(Q)$, to that from the mixed events, $N_{Mix}(Q)$ (filled circles). The ratio of $N_{BEC}(Q)$ to the $Q$ distribution of pure phase space sample, $N_{PS}(Q)$, is also shown (open triangles) in each panel for comparison.  Two solid lines in each pad show the fit results by Eq. (\ref{eqn13082412}) for the ratio $N_{BEC}(Q)/N_{Mix}(Q)$ and $N_{BEC}(Q)/N_{PS}(Q)$, respectively. The corresponding two sets of fit BEC parameters $r_0$, $\lambda_2$ and $N$, and the $\chi^{2}/ndf$ value of the fitting are also shown in each pad.}
\end{center}


\section {Discussion}
Our Monte Carlo simulation results demonstrate that this event mixing method, containing the missing-mass consistency cut and pion-energy cut, is able to identify two boson BEC effects for three body and low multiplicity reactions at low energies. It can make a set of valid reference samples for BEC analysis in such a situation with strict kinematic constraints. The missing-mass consistency cut successfully forces the mixed events in the allowed phase space region. The pion energy cut, found empirically via simulations, is able to produce mixed event samples identical to the original events as long as the event density distribution in terms of $Q$ is concerned. The best pion energy cut value is obtained as a function of the missing mass $m_X$. The reason why the pion energy cut works out is that it eliminates the bias on energy for appropriate partners found in event mixing. 

The simulation aiming to validate this event mixing method shows that it is capable of extracting BEC parameters $r_0$ and $\lambda_2$. As long as the accuracy of the extracted BEC parameters is concerned, this method has a small systematic bias on the obtained $\lambda_2$ and $r_0$, which need to be provided with the corresponding systematic errors. More studies are needed to improve the accuracy of BEC parameters extraction for this event mixing method. 

In the reality, the reaction $\gamma p \to \pi^0\pi^0p$ dose not occur in a manner of pure three body phase space. Many experimental data sets provide evidence that this reaction is dominated by the sequential decay $\gamma p \to \pi^0 \Delta \to \pi^0\pi^0p$ around 1 GeV \cite{Wolf2000, GomezTejedor1996, Thoma2008, Kashevarov2012}. In this situation, special cares may be required in order to make a valid reference sample. The existence of $\Delta$ resonance will significantly affect the kinematical behavior of the three final state particles. As shown in Fig. \ref{fig16031701}, the $\pi^0$ coming from $\Delta$ decay tends to share a relatively smaller momentum compared to the other one. Our event mixing method can take into account such properties to maintain the original statistics. This can be achieved by exchanging two pions both with lower/higher energies from two different events. Because the event mixing method is also sensitive to the pion energy cut, a new appropriate cut value should be found correspondingly. Additionally, the extraction of the dimension parameter $r_0$ becomes more complicated due to the fact that a $\Delta$ appears when the first $\pi^0$ is emitted and then disappears at the same time when the second $\pi^0$ comes out and consequently the time difference between the pair of $\pi^0$s corresponds to the life time of the  $\Delta$. The size of $\Delta$ may be deduced from the space-time interval $r_0$ if one can find a way to eliminate the $\Delta$ decay effect. This special treatment requires further studies.

\begin{center}
\includegraphics[width=0.99\linewidth]{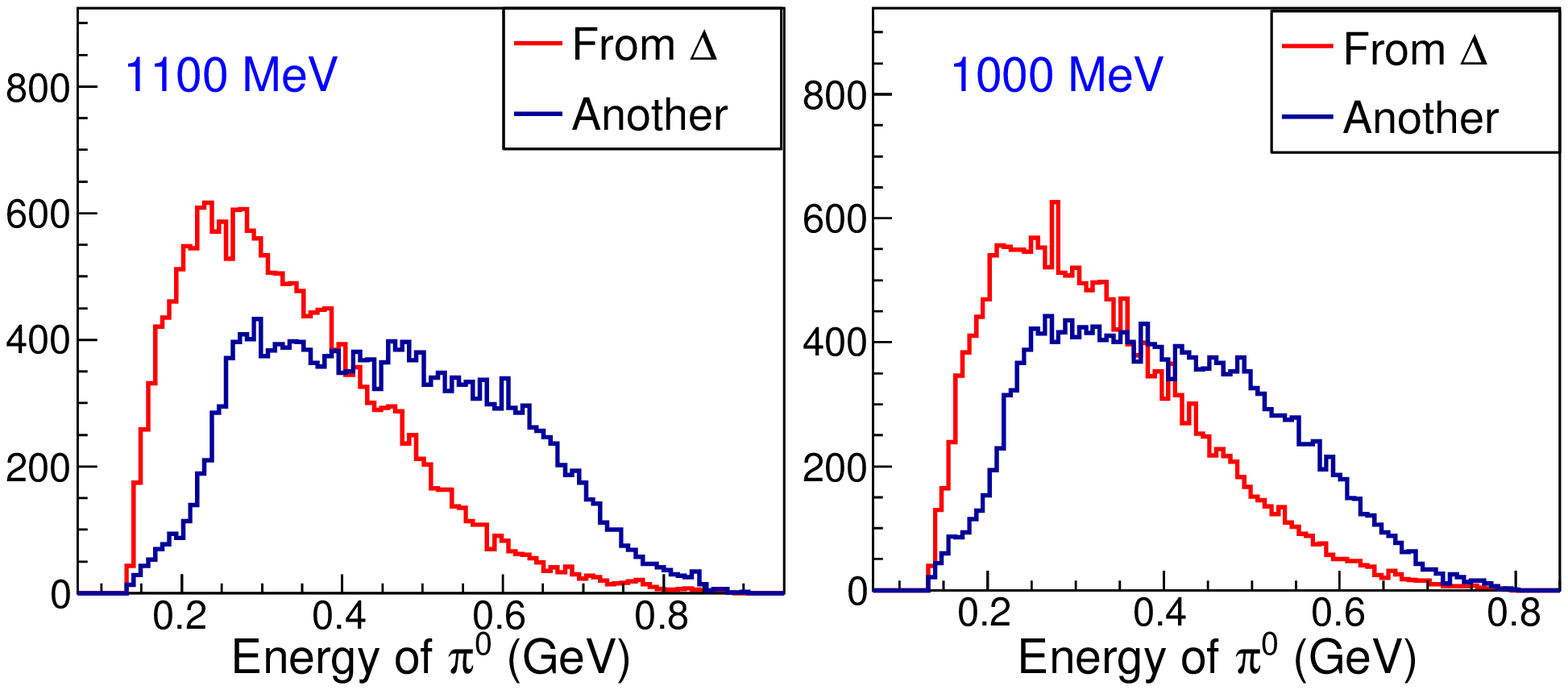}
\figcaption{\label{fig16031701}The spectra of $\pi^0$ energy for the Monte Carlo events $\gamma p \to \pi^0_2 \Delta \to \pi^0_1\pi^0_2p$ at two discrete incident photon energies 1000 MeV and 1100 MeV.}
\end{center}

Although the method is developed on the basis of the two pion photoproduction reaction $\gamma p \rightarrow \pi^{0}  \pi^{0} X$, it is not only applicable for photoproduction, but also similar reactions at lower energies where only two identical bosons are emitted. 

\section {Summary}
We present an event mixing method, which enables to study BEC effects of identical bosons in low multiplicity reactions. This method is equipped with two constraints: (1) missing-mass consistency cut; (2) pion-energy cut. Using several numerical tests for the reaction $\gamma p \rightarrow \pi^{0}  \pi^{0} X$ as an example, the efficiency and correctness of the event mixing method are validated. Applying this method, BEC effects can be observed successfully and corresponding BEC parameters can be extracted.

\section {Acknowledgement}
This work was partially supported by the Ministry of Education and Science of Japan, Grant No. 19002003, and JSPS KAKENHI, Grant No. 24244022 and 26400287.

\end{multicols}

\vspace{-1mm}
\centerline{\rule{80mm}{0.1pt}}
\vspace{2mm}

\begin{multicols}{2}

\end{multicols}

\clearpage
\end{document}